\documentclass[twocolumn,showpacs,preprintnumbers,amsmath,amssymb]{revtex4}
\usepackage{graphicx}
\usepackage{dcolumn}
\usepackage{bm}
\begin{document}

\title{Optical Signatures of Spin-Orbit Interaction Effects in a 
Parabolic Quantum Dot}
\author{Tapash Chakraborty$^{\ast\ddag}$
and Pekka Pietil\"ainen$^{\ast\dag}$}
\affiliation{$^\ast$Department of Physics and Astronomy,
University of Manitoba, Winnipeg, Canada R3T 2N2}
\affiliation{$^\dag$Department of Physical Sciences/Theoretical 
Physics, P.O. Box 3000, FIN-90014 University of Oulu, Finland}
\date{\today}
\begin{abstract}
We demonstrate here that the dipole-allowed optical absorption
spectrum of a parabolic quantum dot subjected to an external 
magnetic field reflects the inter-electron interaction effects 
when the spin-orbit interaction is also taken into account. We have 
investigated the energy spectra and the dipole-allowed transition 
energies for up to four interacting electrons parabolically confined, 
and have uncovered several novel features in those spectra that are 
solely due to the SO interaction.
\end{abstract}
\maketitle

Far-infrared (FIR) optical absorption spectrum of quantum dots,
in particular, dots with a parabolic confinement, has a long and 
colorful history \cite{comment,qdbook,merkt,makchak,farinfrared}. 
First of all, there is a very good theoretical and experimental
understanding of the single-electron states in a quantum dot with 
parabolic confinement. The solution of the Schr\"odinger equation for 
an electron confined by a harmonic potential, $v_c=\frac12m^*\omega_0 
r^2$ in the presence of an external magnetic field, is well
established \cite{qdbook,fock}. The eigenvalues in this case
are given by
$$E_{nl}=\left(2n+\left|l\right|+1\right)\hbar\Omega-\frac12
l\hbar\omega_c$$
where $n,l$ are the principal and azimuthal quantum numbers
respectively, $\Omega^2=\left[\omega_0^2+\frac14\omega_c^2\right]$, 
and $\omega_c$ is the cyclotron frequency. Dipole-allowed transitions 
among these energy levels will have energies \cite{qdbook,merkt,makchak}
$$\Delta E_{\pm}=\hbar\Omega\pm\frac12\hbar\omega_c.$$
This relation has been verified to great accuracy by a variety
of experiments \cite{comment,qdbook,farinfrared}. Interestingly,
however, the observed magnetic field dependent FIR absorption in 
quantum dots with more than one electron was found to be essentially 
{\it independent} of the number of confined electrons and instead 
was dominated by the above relation for $\Delta E_{\pm}$ 
\cite{merkt}. It was a rather
puzzling result because according to this, magneto-optics was
clearly incapable of providing any relevant information about
the effect of mutual interactions of the confined electrons.
The puzzle was later resolved by Maksym and Chakraborty 
\cite{qdbook,makchak}, who pointed out that for a parabolic QD
in an external magnetic field, the dipole interaction is a function of
the center-of-mass (CM) coordinate alone, and only in 
a parabolic confinement the CM terms of the Hamiltonian are separable. 
Since the interaction terms are functions of relative coordinates,
the observed absorption frequencies are independent of the
number of electrons in the dot. Despite this somewhat disappointing
performance of a parabolic dot, FIR spectroscopy of QDs 
(parabolic or otherwise) has generated enormous interest for over 
a decade that is yet to subside \cite{farinfrared}. In this paper 
we demonstrate that, in the presence of spin-orbit coupling the 
situation changes considerably. The energy spectra of
a parabolic quantum dot containing up to four interacting electrons
exhibit structures that are solely due to the presence of SO
interaction. Further, the optical absorption spectra of the spin-orbit
coupled QDs, reported here for the first time, also exhibit novel 
features that are direct reflections of the SO coupling effects.

Interest on the role of the spin-orbit coupling in nanostructured 
systems is now at its peak, due largely to its relevance to spin 
transport in low-dimensional electron channels \cite{spintro}. 
Coherent manipulation of electron spins in low-dimensional systems, 
in particular in quantum dots, is also expected to pave the way for 
future electronic and information processing, especially quantum 
computing and quantum communication \cite{loss}. A recent report 
on the optical detection of spin current in an electron channel 
\cite{spin_Hall} due entirely to the electric field is believed 
to be a major step in that direction. However, most of the experiments 
reported as yet have used magnetotransport measurements \cite{expt}
to gain insight about the spin-orbit coupling effects in
nanostructures. We propose here that magneto-optical 
experiments can also be a very useful alternative in that
pursuit. The spin-orbit interaction that we are here 
concerned with is described by the Hamiltonian 
\cite{rashba,spinorbit}
$${\cal H}_{\rm SO}=
\alpha (\vec k\times\vec\sigma)_z
={\rm i}\alpha \left(\sigma_y\frac{\partial}{\partial
x}-\sigma_x\frac{\partial}{\partial y}\right).$$
Here the $z$ axis is chosen perpendicular to the
2DEG (in the $xy$-plane), $\alpha$ is the spin-orbit
coupling constant, which is sample dependent and is
proportional to the interface electric field,
$\vec{\sigma}=(\sigma_x, \sigma_y, \sigma_z)$
denotes the Pauli spin matrices, and $\vec k$ is the
planar wave vector. In this potential, the spin of
finite-momentum electrons feels a magnetic field
perpendicular to the electron momentum in the
inversion plane. This results in an isotropic spin
splitting energy $\Delta_{\rm SO}$ at $B=0$ that is
proportional to $k$ \cite{spintro,rashba}.

Several experimental groups \cite{expt}
investigating the Shubnikov-de Haas (SdH) oscillations
in a 2DEG confined at the heterojunctions with a
narrow-gap well (e.g., InGaAs/InAlAs, InAs/GaSb, etc.)
have already established that lifting of spin degeneracy
results from inversion asymmetry of the structure which
invokes an electric field perpendicular to the layer.
Experimentally observed values of the SO coupling strength
$\alpha$ lie in the range of 5 - 45 meV nm \cite{expt}.
Energy levels of two interacting electrons confined in
a parabolic quantum dot in an external magnetic field were
recently reported by us for this range of SO coupling strength
\cite{spinorbit}. In the absence of the SO coupling,
electron-electron interaction causes the ground state energy
to jump from one angular momentum value to another as the
magnetic field is increased \cite{qdbook,makchak}. The influence
of the SO coupling is primarily to move the energy level crossings
to weaker fields \cite{spinorbit}. 

The reason why dipole-allowed transitions in a parabolically
confined quantum dot can be very different in the presence of SO
interaction is explained as follows. When subjected to the radiation
field with amplitude $a$ and polarization $\vec\epsilon$,
the vector potential $\vec A$ in the single particle Hamiltonian
\begin{eqnarray*}
{\cal H}_0&=&\frac1{2m}\left(\vec p-\frac ec\vec A\right)^2
+\frac12 m\omega_0^2r^2\\
&&+\frac{\alpha}{\hbar}\left[\vec\sigma\times
\left(\vec p-\frac ec\vec A\right)\right]_z+\frac12 g\mu_BB\sigma_z 
\end{eqnarray*}
must be replaced with the potential
$$ \vec A\rightarrow\vec A+ {\vec A}_\omega, 
 \vec A_\omega=\vec\epsilon a e^{i\vec k\cdot\vec r-i\omega t}.$$
In the dipole approximation we assume that
$$ A_\omega\approx\vec\epsilon a e^{-i\omega t} $$
and correspondingly the Hamiltonian will be \cite{lipparini}
$${\cal H}\approx{\cal H}_0- {\cal H}'\, e^{-i\omega t}, $$
where
$$ {\cal H}'=\frac {ea}{mc}\vec\epsilon\cdot\left(
\vec p-\frac ec\vec A\right)
+\frac{\alpha ea}{\hbar c}\left[\vec\sigma
\times\vec\epsilon\right]_z. $$
In a many-body system when $\alpha=0$ the first term
generates the CM density excitations where mutual interactions
play no role. Consequently (in dipole approximation) only
transitions between these modes are possible. When $\alpha$
is different from zero, the second term ($\propto \sigma_x\epsilon_y
-\sigma_y\epsilon_x$) in ${\cal H}'$ can create spin-density
oscillations and interactions have effect on their properties.  
It is to be noted that, in SO coupled systems
the dipole operator still retains its familiar form,
$\hat Q=\frac{ea}c\vec\epsilon\cdot\vec r$, as is easily verified 
by evaluating its commutator with the Hamiltonian ${\cal H}_0$
$$ [\hat Q,{\cal H}_0]=i\hbar{\cal H}'. $$
Dipole operator is independent of the electron spin. The 
dipole-allowed optical transitions are always from the same 
spin states, but the angular momenta must differ by unity. 
In the presence of SO coupling, neither the dipole operator 
nor the selection rule changes, but the SO interaction
mixes the neighboring angular momentum values ($l$ and 
$l+1$) as well as the spin and hence the selection rule now 
applies to the total angular momentum $J$ as well. Therefore, 
transitions from other states that are not allowed without 
the SO coupling, are now allowed.

\begin{figure}[b]
\begin{center}
 \includegraphics[width=.45\textwidth]{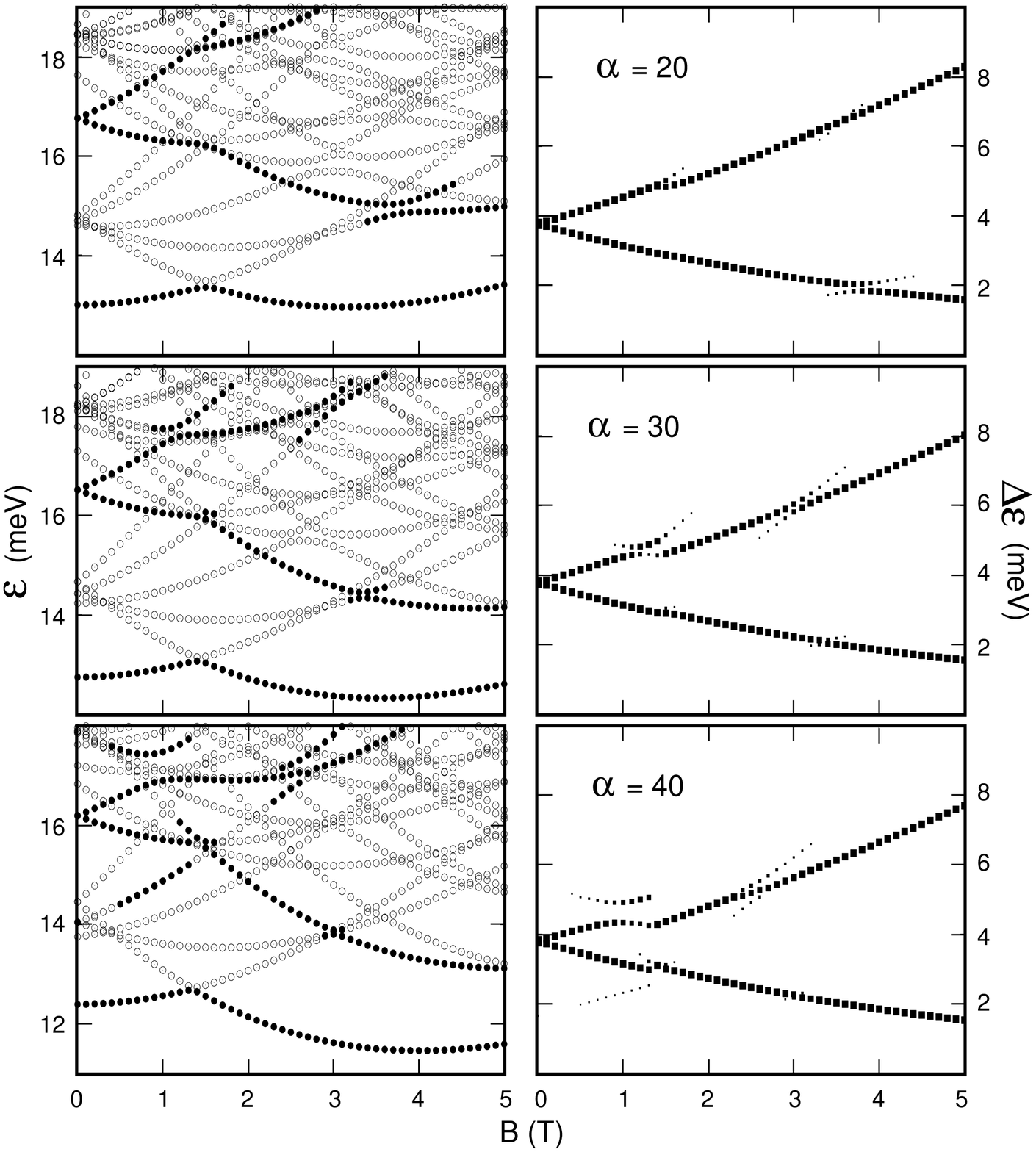}
\protect\caption{
Energy spectra (left panel) and dipole-allowed transition 
energies (right panel) for two interacting electrons confined 
in a InAs quantum dot and for various values of the SO coupling 
strength $\alpha$ [meV nm]. The solid dots in the 
energy spectrum identifies the energy levels involved in transitions 
that correspond to the lowest branches of the absorption spectra (in the
right panel). In the right panels, the size of the points
in the figures is proportional to the calculated intensity.
}\label{fig:dipole1}
\end{center}
\end{figure}

For numerical evaluation of the optical absorption spectrum, we 
have considered a InAs quantum dot where most of the spin-related 
phenomena have been studied \cite{expt}. We have considered up to 
four interacting electrons in the quantum dot. Evaluation of the 
transition energies of a quantum dot containing more than two 
electrons is quite challenging. Firstly, since we work in the 
occupation space spanned by direct anitisymmetrized products of 
the spinors 
$$
|\lambda_l\rangle=
\left(\begin{array}{cc}
\sum_{n=0}u_n^{\lambda_l} f_{nl}e^{il\theta} \\
\sum_{n=0}d_n^{\lambda_l} f_{n,l+1}e^{i(l+1)\theta}
\end{array}\right),
$$
where $f_{nl}=\sqrt{n!/(n+|l|)!}\,x^{|l|/2}e^{-x/2}L_n^l(x)$,
$u_n^{\lambda_l}$ and $d_n^{\lambda_l}$ are the expansion coefficients
\cite{spinorbit}, the two-body matrix elements
are composed of Coulomb matrix elements expressed in a parabolic 
quantum dot as finite sums \cite{qdbook}. In this case however, the 
expansions can extend to Laguerre polynomials of large degree and 
large angular momenta which leads to well known numerical instabilities 
\cite{stone}. To avoid these, we had to resort to multiple precision 
arithmetics (but, in order to circumvent the resulting extremely 
long computation times, we had to tabulate selected subsummations). 
Secondly, since in the presence of the spin-orbit interaction all 
possible spin configurations need to be considered, the dimension 
of the Hilbert space required for convergence can easily become huge 
(of the order of $10^6$ for four electrons). To diagonalize these 
``monster matrices'' we implemented the Davidson-Liu algorithm 
\cite{davidson} that is very suitable for this task. 

\begin{figure}[b]
\begin{center}
 \includegraphics[width=.45\textwidth]{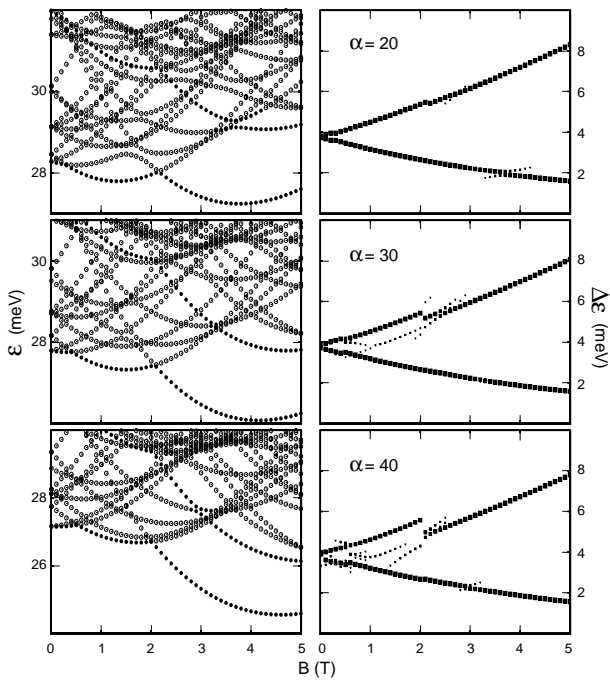}
\protect\caption{Same as in Fig.~1, but for three interacting
electrons in a quantum dot.
}\label{fig:dipole2}
\end{center}
\end{figure}

Our numerical results for energy spectra and absorption
spectra (dipole-allowed) are presented in Figs.~1-3, for
2-4 electrons and for various values of the SO coupling
strength $\alpha$. We have considered the following
parameters for the InAs quantum dot: $m^*/m_0=0.042,
\epsilon=14.6, g=-14$ and $\hbar\omega_0=3.75$ meV.
The dipole matrix elements are evaluated from
\begin{eqnarray*}
d_{\lambda_1\lambda_2}&=&
\langle\lambda_1|re^{i\theta}|\lambda_2\rangle \\
&=&\delta_{l_1,l_2+1}\sum_n\left[
\sqrt{n+l_1}\,u_n^{\lambda_1}u_n^{\lambda_2}
-\sqrt{n}\,u_{n-1}^{\lambda_1}u_n^{\lambda_2}\right. \\
&&\left. +\sqrt{n+l_1+1}\,d_n^{\lambda_1}d_n^{\lambda_2}
-\sqrt{n}\,d_{n-1}^{\lambda_1}d_n^{\lambda_2}\right]
\end{eqnarray*}
when $l_1\geq0$ and a similar type of equation for $l_1<0$.
The intensity is obtained from
${\cal I}\propto \vert d_{\lambda_1\lambda_2}\vert^2$ \cite{halonen}.
In our plots for the absorption spectra (the right panels of the figures),
the size of the points is proportional to the calculated intensity.

\begin{figure}[b]
\begin{center}
 \includegraphics[width=.45\textwidth]{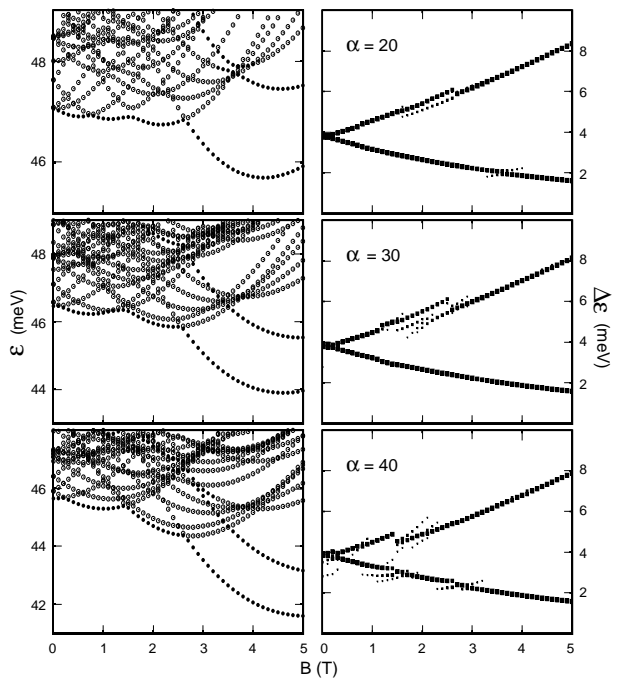}
\protect\caption{Same as in Fig.~1, but for four interacting
electrons in a quantum dot.
}\label{fig:dipole3}
\end{center}
\end{figure}

A striking feature visible in the absorption spectra (right hand 
panels of the figures) is the appearance of discontinuities, 
anticrossings and new modes in addition to the two main 
($\alpha=0$) absorption lines. These optical signatures of 
the SO interaction are consequences of the multitude of 
level crossings and level repulsions that occur in the 
energy spectra (left hand panels of Figs.~1-3). The latter 
ones can be attributed to an interplay between SO and Zeeman 
couplings. In order to understand their origin, let us first examine
the case of the two electron system. In our spinor notation the main 
contribution to the ground state at zero magnetic field comes from the 
two-electron state
$|\lambda_{l_1},\lambda_{l_2}\rangle=|\lambda_0,\lambda_{-1}\rangle$,
where $|\lambda_{l_1}\rangle$ is a spinor with $J_1=l_1+1/2=1/2$,
$d_n^{\lambda_1}=0$, and $|\lambda_{l_2}\rangle$
a spinor with $J_2=-1/2$ and $u_n^{\lambda_2}=0$,
i.e., both electrons have zero orbital angular momenta with opposite spins
(corresponding to $J=J_1+J_2=0$). Now when we increase the magnetic field
the spin triplet configuration will become, due to the interaction,
energetically more favorable. If the Lande factor is negative then
the electrons would like to occupy states with orbital angular momenta
0 and $-1$ with both spins up. In the spinor picture this means that
$|\lambda_{l_2}\rangle$ still has $l_2=-1$ ($J=0$)
but now $u_n^{\lambda_2}\not=0$ and $d_n^{\lambda_2}=0$.
The SO interaction mixes these two configurations which results in
a level repulsion. On the other hand, when the strength of the
SO coupling is further increased, the relative significance of the Zeeman
contribution to ${\cal H}_0$ decreases. The energy shifts to states with 
$J\not=0$ will then become energetically feasible and we have again 
crossings of levels. For increasing number of electrons in the dot, the
energy spectra is more dense and exhibit additional level crossings [Figs.~2-3].
As a consequence, the ground state momentum also changes more frequently
as compared to that of the two electron case.

At moderate SO coupling strengths the absorption spectra do not 
essentially differ from the single particle spectrum.  But when the coupling 
strength increases the deviation from the pure parabolic confinement also 
increases which in turn implies that the lowest final states
of dipole allowed transitions are not any more achievable by adding
$\hbar\Omega\pm\frac12\hbar\omega_c$ to the initial state energies.
In particular, this results in discontinuities and anticrossing
behaviors as well as appearance of new modes. As an illustration,
let us consider the absorptions that at the magnetic field
$B=1$T take the two electron system from the ground state to excited
states. In the absence of the SO coupling the ground state is a
spin singlet state $S =0$ with total angular momentum $J=0$.
According to the dipole selection rules absorptions cause transitions to
states $J=\pm1$ and $S=0$ with energies $\Delta E_\pm$ above
the ground state. Looking at the bottom right panel of Fig.~1,
we note that in addition to the two main lines there are now two additional
lines (at $B=1$ T) of appreciable intensity at the SO coupling strength
$\alpha=40$.
Further analysis reveals that the ground states still have $J=0$ and that
the expectation value of the spin $z$-component is $\langle\sigma_z\rangle=0$.
The excited states also have $J=\pm1$, as before. However, the final
spin states can no longer be classified as singlets: the expectation
values $\langle\sigma_z\rangle$ vary between $-0.03$ and 0.39.
When the number of electrons increases the number of these additional
modes also increases but at the same time the relative intensities
decrease (at each $B$ we have normalized the total intensity to unity).
On the other hand, the discontinuities as consequences of deviations
from a parabolic confinement become more pronounced (Figs.~2-3).

In conclusion, we have presented the energy levels and optical 
absorption spectra of up to four interacting electrons in a 
parabolic quantum dot subjected to an externally applied perpendicular 
magnetic field. The optical absorption spectra exhibit new modes
that are a direct consequence of the SO coupling effects. These are
manifested in the energy spectra as multiple level crossings and level 
repulsions that are attributed to an interplay between the Zeeman and 
SO couplings present in the system Hamiltonian. 

We would like to thank Marco Califano for many helpful discussions. 
The work of T.C. has been supported by the Canada Research Chair 
Program and the Canadian Foundation for Innovation (CFI) Grant.

\end{document}